\documentclass[letter]{aa} 
\usepackage{graphicx}
\usepackage{txfonts}
\usepackage{aalongtable}
\begin{document}
\titlerunning{Short-term variability on the surface of (1) Ceres}
\authorrunning{D. Perna et al.} 

   \title{Short-term variability on the surface of (1) Ceres
\thanks{Based on observations made with the Italian Telescopio Nazionale Galileo (TNG)
operated on the island of La Palma by the Fundaci\'{o}n Galileo Galilei of the INAF
(Istituto Nazionale di Astrofisica) at the Spanish Observatorio del Roque de los Muchachos
of the Instituto de Astrofisica de Canarias.}}

   \subtitle{A changing amount of water ice?}

\author{D.~Perna\inst{1}
\and
Z.~Ka\v{n}uchov\'{a}\inst{2}
\and
S.~Ieva\inst{1,3,4}
\and
S.~Fornasier\inst{1,5}
\and
M. A. Barucci \inst{1}
\and
C.~Lantz\inst{1,5}
\and
E. Dotto \inst{3}
\and
G.~Strazzulla\inst{6}
}

\institute{LESIA -- Observatoire de Paris, CNRS, UPMC Univ. Paris 06, Univ. Paris-Diderot, 5 place J. Janssen, 92195, Meudon, France\\
              \email{davide.perna@obspm.fr}
\and
Astronomical Institute of the Slovak Academy of Sciences, 059 60 Tatransk\'{a} Lomnica, Slovakia
\and
INAF -- Osservatorio Astronomico di Roma, Via Frascati 33, 00040 Monte Porzio Catone (Roma), Italy
\and
Universit\`a di Roma Tor Vergata, Via della Ricerca Scientifica 1, 00133 Roma, Italy
\and
Universit\'{e} Paris Diderot -- Paris 7, 4 rue Elsa Morante, 75013 Paris, France
\and
INAF -- Osservatorio Astrofisico di Catania, Via S. Sofia 78, 95123 Catania, Italy
}

   \date{Received 9 November 2014 / Accepted 24 January 2015}

 
  \abstract
   {The dwarf planet (1) Ceres -- next target of the NASA Dawn mission -- is the largest body in the asteroid main belt.
Although several observations of this body have been performed so far,
the presence of surface water ice is still questioned.}
   {Our goal is to better understand the surface composition of Ceres,
and to constrain the presence of exposed water ice.}
   {We acquired new visible and near-infrared spectra at the Telescopio Nazionale Galileo (TNG, La Palma, Spain),
and reanalyzed literature spectra in the 3-$\mu$m region.}
   {We obtained the first rotationally-resolved spectroscopic observations of Ceres at visible wavelengths.
Visible spectra taken one month apart at almost the same planetocentric coordinates
show a significant slope variation (up to 3 \%/10$^3\AA$).
A faint absorption centered at 0.67 $\mu$m, possibly due to aqueous alteration, is detected in a subset of our spectra.
The various explanations in the literature for the 3.06-$\mu$m feature can be interpreted
as due to a variable amount of surface water ice at different epochs.}
   {The remarkable short-term temporal variability of the visible spectral slope,
and the changing shape of the 3.06-$\mu$m band,
can be hints of different amounts of water ice exposed on the surface of Ceres.
This would be in agreement with the recent detection by the Herschel Space Observatory
of localized and transient sources of water vapour over this dwarf planet.}


   \maketitle
%

\section{Introduction}
The dwarf planet (1) Ceres is the only one in the inner solar system.
It probably represents a unique remnant of the primordial population
of large planetesimals that has been later on removed by dynamical mechanisms,
depleting the asteroid main belt of most of its original mass (e.g., Morbidelli et al. 2009).
The modeling of the thermo-physico-chemical evolution of Ceres,
which is constrained by the available shape and density measurements
(2077$\pm$36 kg m$^{-3}$, Thomas et al. 2005; 2206$\pm$43 kg m$^{-3}$, Carry et al. 2008),
points to a differentiated interior. This probably consists of an inner core
made of dry silicates, an outer layer made of hydrated silicates, and
an outer shell of water ice (Castillo-Rogez and McCord 2010).
Alternatively, an undifferentiated porous interior, consisting of a mixture of
hydrated silicates, has been proposed (Zolotov 2009).
The albedo (0.090$\pm$0.003, Li et al. 2006) as well as the visible and near-infrared spectra of Ceres
are similar to those of carbonaceous chondrites, and evidence for aqueous alteration has been found.
The presence of surface water ice on Ceres,
possibly resupplied from the mantle through fractures in a silicate and ice crust, has been debated at length
(see Rivkin et al. 2011 for a review about Ceres' surface composition).
So far, rotationally-resolved spectra of Ceres (P = 9.074170$\pm$0.000002 h, Chamberlain et al. 2007)
were acquired in the near-infrared region
by Rivkin and Volquardsen (2010) and Carry et al. (2012), who reported evidence
for a relatively uniform surface.
Before the present study, no study of visible spectra variation with the Ceres' rotation had been performed.
Recently, K\"uppers et al. (2014) identified two localized sources ($\sim$60 km in diameter) of
water vapour on Ceres, at planetocentric longitudes corresponding to
the dark regions Piazzi (long. 123\degr, lat. +21\degr) and Region A (long. 231\degr, lat. +23\degr).
The water outgassing has been suggested to be due to comet-like sublimation or to cryo-volcanism.

In this work we analyze new visible and near-infrared spectra of Ceres
obtained at different rotational phases,
in order to enhance our knowledge of its surface composition.
We also review the interpretation of the 3-$\mu$m feature observed in Ceres' spectra by different authors.
Our results can be useful in support
of the NASA Dawn mission (Russell et al. 2004), that
will start orbiting Ceres in March 2015.

\section{Observations and data reduction}
\begin{table*}[t]
\caption{Observational circumstances and spectral parameters. SEP$_{\lambda}$ and SEP$_{\beta}$ are the sub-Earth point longitude and latitude.} 
\label{table:1}      
\centering
\scriptsize{
\begin{tabular}{ccccccccccc}        
\hline\hline                 
ID & Date 	 & UT    & Prism/Grism & t$_{exp}$ (s) & Airmass & Solar analog (airm.) & SEP$_{\lambda}$ & SEP$_{\beta}$ & Vis. Slope (\%/10$^3\AA$) & 0.67 $\mu$m band \\
\hline                        
A  & 2012 Dec 17 & 22:10 & Amici       & 4$\times$3  &1.31     & Land115-271 (1.27)     & 74.1            & -3.0          & -- & -- \\
B  & 2012 Dec 18 & 01:05 & Amici       & 4$\times$3  &1.00     & Hip9197 (1.01)         & 318.3           & -3.1          & -- & -- \\
\hline                        
1   & 2012 Dec 18 & 02:31 & LR-B+LR-R & 1+1         & 1.05   & Hip599932 (1.06)       & 261.1           & -3.1          & -1.15$\pm$0.29 & 0.6--0.7\% \\
2   & 2012 Dec 18 & 23:27 & LR-B+LR-R & 1+1         & 1.08   & Hyades142 (1.05)       & 150.5           & -3.1          & -0.70$\pm$0.23 & 0.5--0.6\% \\
3   & 2012 Dec 19 & 02:34 & LR-B+LR-R & 1+1         & 1.06     &Hyades142 (1.05)      & 26.8            & -3.2          & -3.18$\pm$0.62 & $<$0.5\% \\
4   & 2013 Jan 18 & 20:12 & LR-B+LR-R & 1+1         & 1.18     &Hip44027 (1.18)       & 277.3           & -5.6          & +1.26$\pm$0.36 & $<$0.5\% \\
5   & 2013 Jan 19 & 20:44 & LR-B+LR-R & 1+1         & 1.09     &Hyades64 (1.03)       & 23.9            & -5.7          & +0.05$\pm$0.42 & $<$0.5\% \\
6   & 2013 Jan 20 & 01:43 & LR-B+LR-R & 3+5         & 1.35     &Land102-1081 (1.58)   & 186.2           & -5.7          & +0.01$\pm$0.24 & $<$0.5\% \\
\hline
\end{tabular}}
\end{table*}

Observations were carried out at the 3.6-m Telescopio Nazionale Galileo (La Palma, Spain),
during two runs (December 2012 and January 2013).
The observational circumstances are given in Table~1.
Visible spectroscopy was performed with the DOLORES instrument, using the low resolution blue (LR-B)
and red (LR-R) grisms.
Near-infrared (NIR) data were taken with the NICS instrument coupled with the Amici prism, in low-
resolution mode. NIR observations were carried out using the standard nodding technique of moving the object along the slit
between two positions A and B. Each of the two NICS spectra that we present was obtained from one ABBA sequence.
For both visible and NIR ranges, spectra were taken through a 2\arcsec slit, oriented along the parallactic angle to
avoid flux loss due to the differential refraction.

Data reduction was performed with the software package
ESO-Midas, using standard procedures (e.g., Fornasier et al. 2004). Namely, for the visible data:
subtraction of the bias from the raw data, flat field correction, cosmic rays
removal, background subtraction, collapsing the two--dimensional spectra into one dimension,
wavelength calibration (using Ar, Ne+Hg, and Kr lamps' emission lines) and atmospheric extinction correction.
NIR data were corrected for flat field, A-B and B-A pairs were subtracted,
the four different frames were shifted and added, then the spectrum was extracted.
Wavelength calibration was obtained using a look-up table,
available on the instrument website, which is based on the theoretical dispersion
predicted by ray-tracing and adjusted to best fit the observed spectra of
calibration sources.
The reflectivity of Ceres was then obtained by dividing its visible and NIR spectra by those of
solar analogs observed close in time and in airmass to the scientific frames (cf. Table~1).

\begin{figure}
   \centering
   \includegraphics[angle=0,width=9cm]{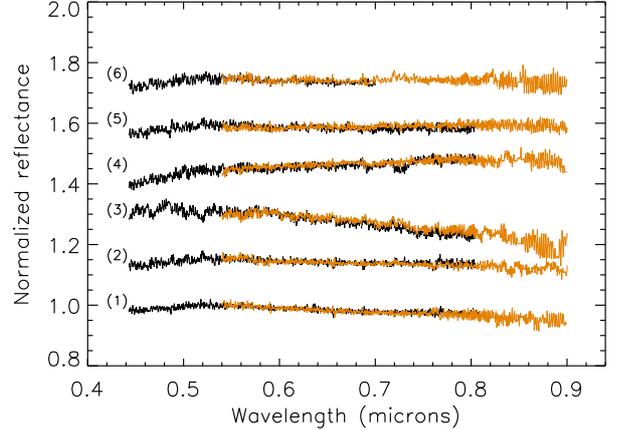}
      \caption{Visible LR-B (black) and LR-R (orange) spectra of Ceres,
shifted by 0.15 in reflectance for clarity. The LR-B spectrum \#6 has been cut at 0.7 $\mu$m because
affected by a poor sky subtraction longward.}
         \label{fig1}
   \end{figure}

\begin{figure}
   \centering
   \includegraphics[angle=0,width=9cm]{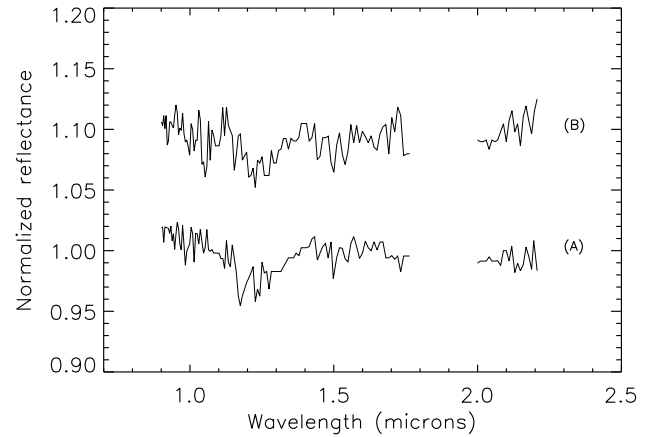}
      \caption{NIR spectra of Ceres, shifted by 0.1 in reflectance for clarity.
Regions affected by strong atmospheric absorption have been cut out.}
         \label{fig2}
   \end{figure}

The resulting visible and NIR spectra are presented in Fig.~\ref{fig1} and Fig.~\ref{fig2}, respectively.
During each of the two observing runs, we obtained 3 complete spectra of Ceres with the LR-B + LR-R grisms
(covering the 0.44--0.81 $\mu$m and 0.54--0.9 $\mu$m wavelength ranges, respectively).
The planetocentric coordinates for each observation (Table~1) have been calculated using the IMCCE webtools (Berthier et al. 2008),
according to the pole solution by Carry et al. (2008).
Two NIR spectra have been also acquired during the December 2012 run.

\section{Data analysis and results}
The two NIR spectra we obtained do not show significant variations, in agreement with Carry et al. (2012).
Though noisy and of low resolution, both of them show the very broad absorption centered at $\sim$1.2 $\mu$m
already observed on Ceres, as well as on several carbonaceous chondrites, and interpreted
as possibly due to magnetite or plagioclase feldspar (Rivkin et al. 2011).

\begin{figure}
   \centering
   \includegraphics[angle=0,width=9cm]{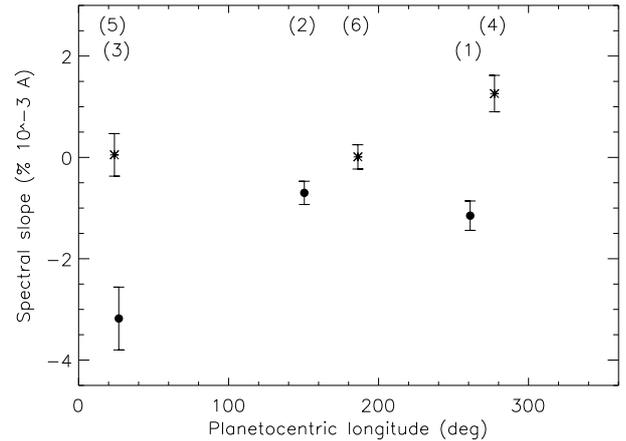}
      \caption{Visible spectral slope (0.55--0.8 $\mu$m) vs. planetocentric longitude.
Data from Dec. 2012 (dots) and Jan. 2013 (asterisks).}
         \label{fig3}
   \end{figure}

The visible spectra obtained from the 6 simultaneous LR-B and LR-R observations
show a very good agreement between the two grisms, making us confident about the goodness of our results.
In Table~1 we report the spectral slope computed in the 0.55-0.8 $\mu$m range,  
considering together the LR-B and LR-R data.
Fig.~\ref{fig3} shows the obtained spectral slopes as a function
of the longitude of the sub-observer point.
A variation through the rotational phase is evident,
whose amplitude is about one order of magnitude larger than that reported for NIR observations (Carry et al. 2012).
We also stress that, while all of the spectra taken in January 2013 (phase angle $\sim$13.5\degr) are redder than
those acquired in December 2012 (phase angle $\sim$0.7\degr), the observed variations are about
two orders of magnitude larger than those that could be introduced by the phase reddening effect (e.g., Sanchez et al. 2012).

While we cannot determine the exact cause of these variations (possibly due e.g. to different composition,
albedo, geological features, surface age), we stress that the spectra \#3 and \#5 -- as well as, to a minor extent, \#1 and \#4 --
sample the same regions on the surface, observed one month apart,
and present a significant variation of the slope ($\sim$2--3 \%/10$^3\AA$).
Noteworthy, our observing runs (December 2012 and January 2013) are temporally encompassed
by the water vapour detections around Ceres (October 2012 and March 2013)
performed with the Herschel Space Observatory by K\"uppers et al. (2014).
Hence we guess that the observed variation of the visible spectral slope could be
linked with the resurfacing after outgassing episodes.
The bluer spectrum could correspond to larger quantities of exposed water ice
and/or some ``B-type'' asteroidal material (possibly exposed and/or redeposited by water sublimation).
The redder spectrum could instead be tentatively associated with more exposed organics:
though these have not been yet clearly detected, their presence
on Ceres -- as for water ice -- is considered likely based on thermo-physical considerations (Castillo-Rogez and McCord 2010)
and they could possibly lie in the near-surface environment (Rivkin et al. 2011).
Moreover, a mix of water ice and organics has already been invoked to fit the Ceres' 3-$\mu$m band (see next section).
The much smaller slope variations observed by Carry et al. (2012) in the NIR
could then be associated with periods of no/lower water activity by Ceres
(e.g., Herschel did not detect the water vapour on November 2011);
or the Ceres' spectrum could be intrinsically less variable at NIR wavelengths than
in the visible.
Our spectra \#2 and \#6, acquired at planetocentric longitudes $\sim$30\degr apart,
show a slope variation similar to that previously reported for the NIR.
This could strengthen the hypothesis of localized active regions on the surface of Ceres.

\begin{figure}
   \centering
   \includegraphics[angle=0,width=7cm]{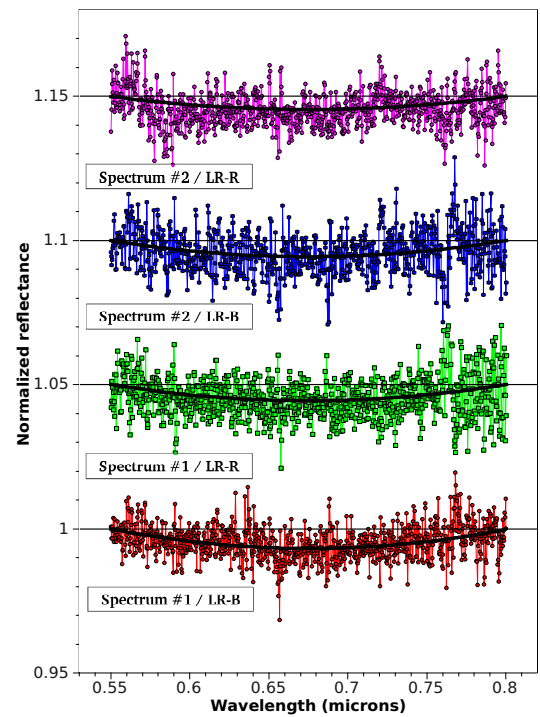}
      \caption{Continuum-removed LR-B and LR-R spectra \#1 and \#2 (shifted by 0.05 in reflectance for clarity),
with a superimposed polynomial fit.}
         \label{fig4}
   \end{figure}

The existence and spatial extent on Ceres 
of a broad spectral feature around 0.6--0.7 $\mu$m,
usually attributed to aqueous alteration products
(e.g. saponite, Cloutis et al. 2011),
is still open to debate (Rivkin et al. 2011).
To verify the presence and variability of such absorption
in our spectra, we removed the continuum computed as a linear fit to
each spectrum in the region 0.55--0.8 $\mu$m.
A very faint feature (0.5--0.7\%) centered at $\sim$0.67 $\mu$m seems present in
spectra \#1 and \#2, for both LR-B and LR-R grisms;
no noticeable absorption is identified in our further spectra,
maybe just because of lower S/N (Table~1 and Fig.~\ref{fig4}).

For comparison, we also reanalyzed the few visible spectra of Ceres available in the literature,
computing the spectral slope and 
looking for the presence of the aqueous alteration band (Table~2):
an absorption centered at 0.67 $\mu$m and with depth 1.7\% is clearly seen in the spectrum by Vilas et al. (1993);
the spectrum from Vilas \& McFadden (1992) seems to exhibit two superimposed features with minima at 0.6 $\mu$m and 0.67 $\mu$m,
with an overall depth $<$1.0\%, though the double minimum could be an artifact attributable to poor S/N;
a similar result is found for the spectrum by Fornasier et al. (1999), while no feature is seen in the data
by Bus \& Binzel (2002) and Lazzaro et al. (2004).
As for the spectral slope, we note that
the magnitude of its variation is less considerable in literature data than in our dataset,
despite of the uncertainties due to the different observing conditions and techniques, used instrumentation, etcetera.

\begin{table}[t]
\caption{Spectral parameters for literature visible data.}             
\label{table:2}      
\resizebox{\columnwidth}{!}{
\begin{tabular}{lccc}        
\hline\hline                 
Reference & Date 	 & Slope (\%/10$^3\AA$) & 0.67 $\mu$m band \\
\hline                        
Vilas \& McFadden (1992) & 4 Jul 1987 &  -0.76$\pm$0.27& ? ($<$1.0\%) \\
Vilas et al. (1993) & 20 Apr 1992 & 0.25$\pm$0.66 & Y (1.7\%)\\
Bus \& Binzel (2002) & 1993--1998 & 0.72$\pm$1.00 & N ($<$0.3\%) \\
Lazzaro et al. (2004) & 13 Jul 1997 & 0.95$\pm$0.19 & N ($<$0.3\%) \\
Fornasier et al. (1999) & 13 Dec 1997 & -0.54$\pm$0.36 & ? ($<$1.0\%) \\ 
\hline
\end{tabular}}
\end{table}

\section{The 3-$\mu$m band and the water ice on Ceres} \label{sec:3micron}
A broad 3-$\mu$m feature was observed on Ceres for the first time by Lebofsky (1978),
and associated with the presence of clay minerals containing water of hydratation.
Analyzing higher spectral resolution data, the 3-$\mu$m feature was confirmed by Lebofsky et al. (1981),
who also found a relatively narrow absorption centered near 3.06 $\mu$m, interpreted as due
to a very thin layer of water ice frost.
This feature, along with a set of overlapping bands at 3.3--3.4 $\mu$m, was reinterpreted by King et al. (1992)
as due to ammoniated phyllosilicates.
Vernazza et al. (2005) revisited ice as an explanation, reproducing the 3.06-$\mu$m feature with
a linear combination of crystalline water ice and an organic component, namely
residues of ion-irradiated asphaltite covered with water ice. 
Rivkin et al. (2006) further reinterpreted the 3.06-$\mu$m band as due to an iron-rich clay (the serpentine cronstedtite) and
the 3.3--3.4 and 3.8--3.9 $\mu$m bands as caused by carbonates; the same dataset was then modeled by Milliken and Rivkin (2009)
to identify the 3.06-$\mu$m band as brucite (Mg(OH)$_{2}$) and confirm the presence of carbonates.
Beck et al. (2011) proposed impure goethite (FeO(OH)) as an alternative origin of the 3.06-$\mu$m band,
though Jewitt and Guilbert-Lepoutre (2012) stressed that goethite, when found in meteorites, is a product of
aqueous alteration in the terrestrial environment, and that extraterrestrial goethite in freshly fallen
meteorites is unknown.

\begin{figure}
   \centering
   \includegraphics[angle=0,width=8cm]{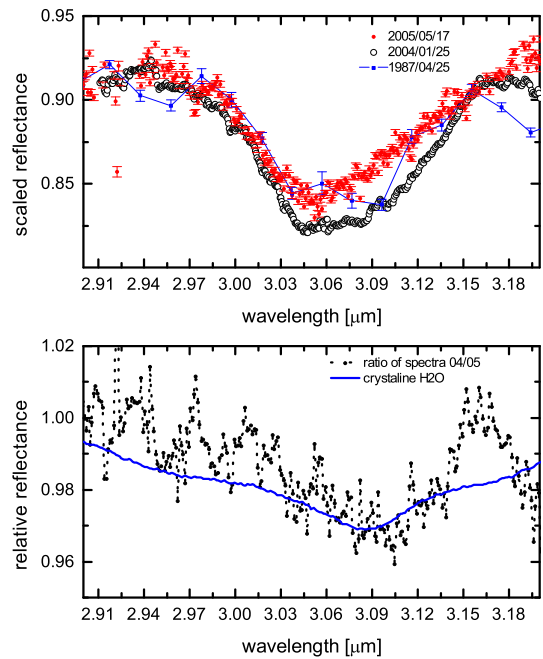}
      \caption{Up: Ceres' 3-$\mu$m spectra taken in 1987 (Jones et al. 1990), 2004 (Vernazza et al. 2005), 2005 (Rivkin et al. 2006).
Down: the 2004/2005 spectral ratio and the spectrum of crystalline (150 K) water ice.}
         \label{fig5}
   \end{figure}

In Fig.~\ref{fig5} we show the 3-$\mu$m region spectra by
Jones et al. (1990), Vernazza et al. (2005), and Rivkin et al. (2006),
acquired in 1987, 2004, and 2005, respectively.
In the first two works, the presence of the 3.1-$\mu$m absorption due to crystalline water was claimed. 
In Fig.~\ref{fig5} we also show the ratio of the high resolution 2004/2005 spectra, as well as 
a laboratory spectrum of crystalline (150 K) water ice (Strazzulla et al. 2001).
Supposing that the 3-$\mu$m-band is composed of two components, 
as suggested by Vernazza et al. (2005), the ratio of the two spectra
may reveal a change in the relative abundances.
Indeed, the ratio of the spectra from 2004 and 2005 is very similar to the spectrum of crystalline water, with a minimum
at $\sim$3.1 $\mu$m (the absorption starting at 3.15 $\mu$m in the 2004 spectrum is due to the terrestrial atmosphere,
as stated by Vernazza et al. 2005).

Following our findings about the short-term temporal variation of the visible spectral slope,
as well as the recent identification of localized emission of water vapour around Ceres,
we suggest that also the shape of the 3-$\mu$m band could change in time:
i.e., Vernazza et al. (2005)
could have really observed crystalline water ice, while Rivkin et al. (2006)
could have observed Ceres when surface ice was less abundant and not detectable.

\section{Summary and conclusions}
%
The new visible and NIR spectra that we analyze in this work are overall compatible with previous data of Ceres;
the feature centered at 0.67 $\mu$m seen in previous observations (e.g., Vilas et al. 1993), and interpreted as due to aqueous alteration,
is possibly observed in two out of our six visible spectra (those with higher S/N),
although the measured band depth is very small (0.5--0.7 \%).
Data taken in December 2012 and January 2013 show a clear variation of the visible spectral slope:
at longitudes $\sim$25$\degr$ and $\sim$270$\degr$ we measure a slope difference of
(3.2$\pm$1.0)\%/10$^3\AA$ and (2.4$\pm$0.7)\%/10$^3\AA$, respectively.
Such variations are much more important than those already reported in the literature and strongly point towards
a variable character of the Ceres' surface.
While the regions we identified
(which are however approximate due to our disk-integrated observations)
do not correspond with those detected as emitting water vapour
by K\"uppers et al. (2014) in October 2012 and January 2013,
we can hypothesize that Ceres was undergoing extended resurfacing processes (due to cryovolcanism or cometary activity) during that period.  
These new results also lead us to propose a coherent and reconciling solution for the disagreements in the literature about the
shape and interpretation of the 3-$\mu$m band, 
that could change in time according to the different quantity of exposed water ice.
The NASA Dawn mission that will start to orbit Ceres in March 2015 will hopefully provide
key information about the surface variegation and any outgassing activity of Ceres.

\begin{acknowledgements}
We thank A.S. Rivkin and P. Vernazza for kindly providing us their spectral data.
The research of Z.K. is supported by VEGA -- The Slovack Agency for Science, Grant No. 2/0032/14.
This program was supported by the French national space agency CNES and French INSU -- PNP program.

\end{acknowledgements}

\Online

\begin{appendix} 

\section{About the solar analog stars}
\begin{figure}
\centering
\includegraphics[angle=0,width=9cm]{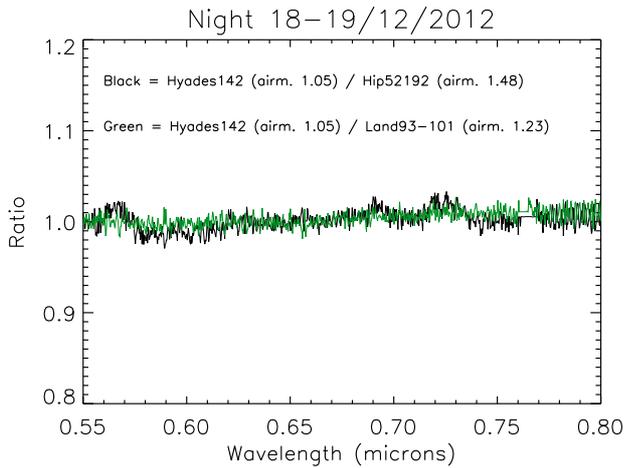}
\caption{Spectral ratios of the three solar analog stars observed during the night of 18-19 Dec. 2012.
The airmasses at the moment of the acquisition are reported.}\label{ratio2}
\end{figure}

\begin{figure}
\centering
\includegraphics[angle=0,width=9cm]{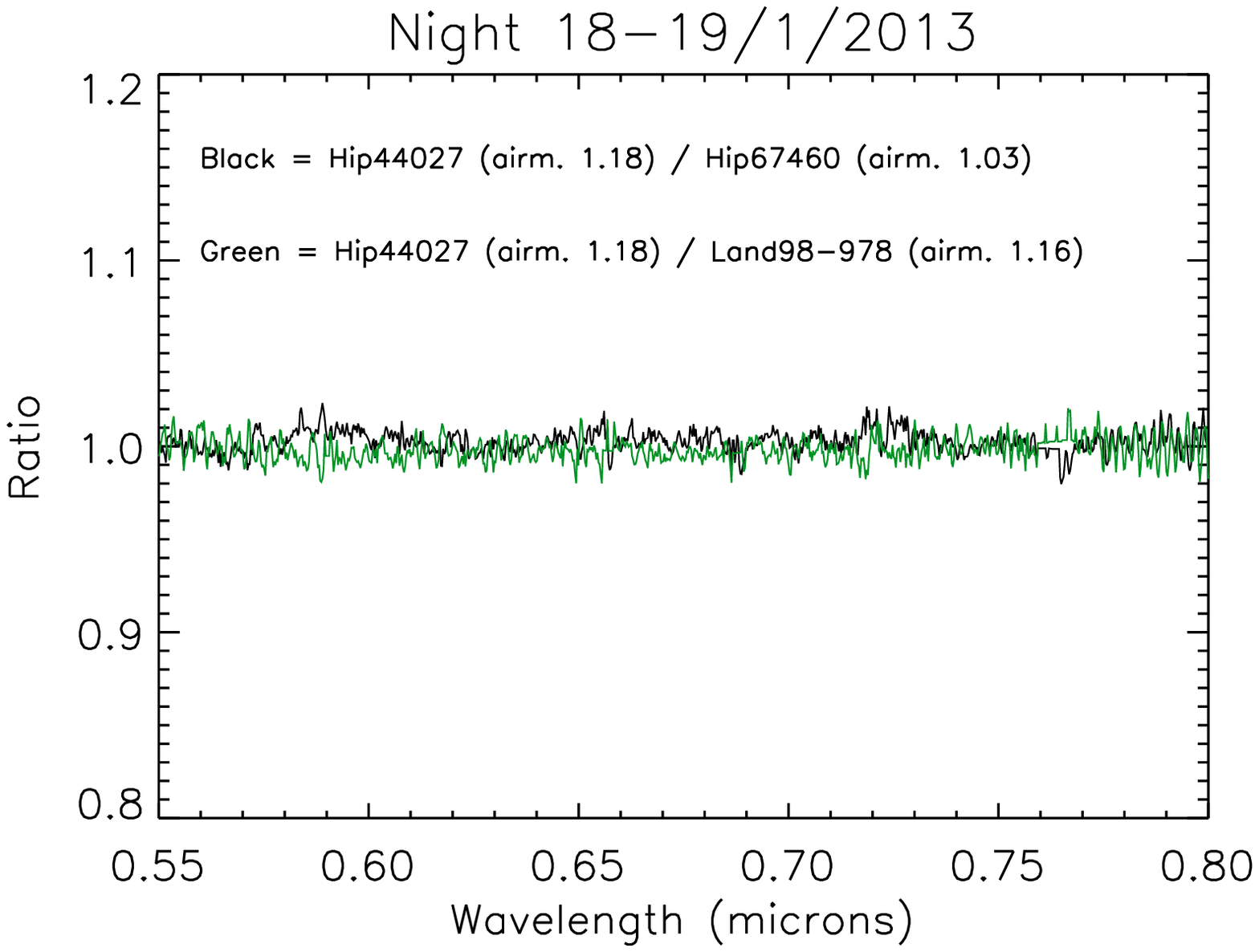}
\caption{Spectral ratios of the three solar analog stars observed during the night of 18-19 Jan. 2013.
The airmasses at the moment of the acquisition are reported.}\label{ratio3}
\end{figure}

\begin{figure}
\centering
\includegraphics[angle=0,width=9cm]{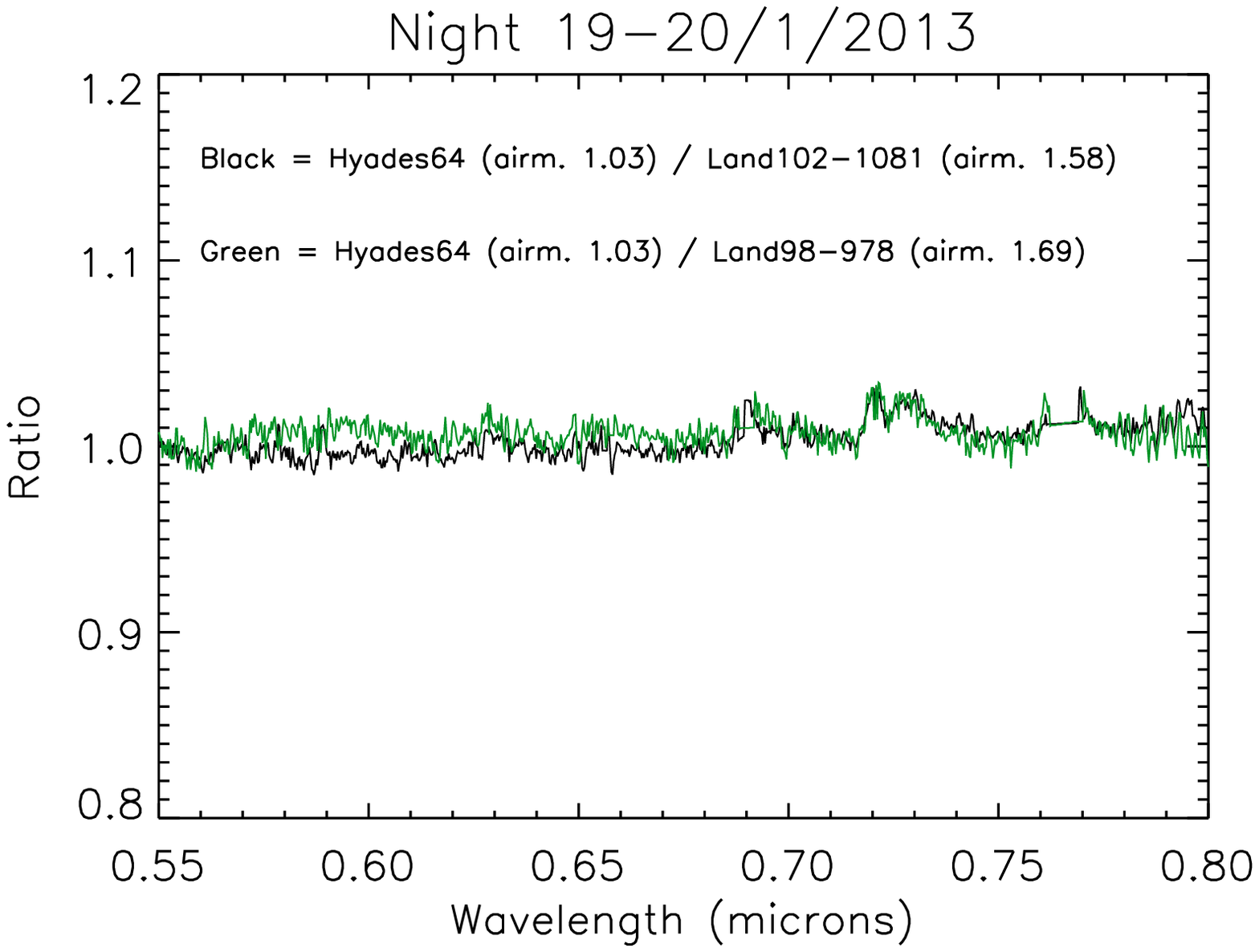}
\caption{Spectral ratios of the three solar analog stars observed during the night of 19-20 Jan. 2013.
The airmasses at the moment of the acquisition are reported.}\label{ratio4}
\end{figure}

\begin{figure}
\centering
\includegraphics[angle=0,width=9cm]{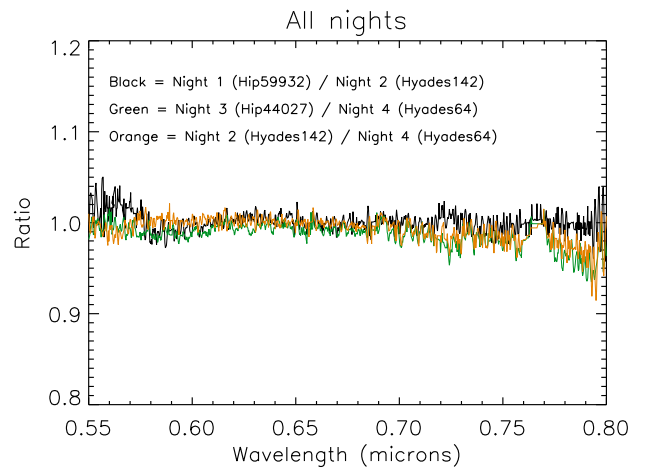}
\caption{Spectral ratios of solar analog stars observed during different nights.
Night 1 = 17-18 Dec. 2012,
Night 2 = 18-19 Dec. 2012,
Night 3 = 18-19 Jan. 2013,
Night 4 = 19-20 Dec. 2013.}\label{ratio_all}
\end{figure}

To verify if the observed variability of Ceres' visible spectra is real,
and not an artifact due to the different solar analog stars used for calibration,
we divided them by one another to check if this results in a flat spectrum
(we considered the LR-R spectra in the same wavelength range used for computing
the Ceres' visible spectral slope, 0.55--0.8 $\mu$m).
Figures \ref{ratio2}~to~\ref{ratio4} show the relative ratios of the three solar analog stars
observed during each night, on 18-19/12/2012, 18-19/1/2013 and 19-20/1/2013, respectively.
For the second of these nights (Fig.~\ref{ratio3}), the stars were observed at low airmass and their spectral ratios
are flat (slopes of -0.02$\pm$0.15\%/10$^3\AA$ and 0.10$\pm$0.18\%/10$^3\AA$);
The spectral ratios of stars observed during the other two nights
present a slightly positive slope ($\la$0.5\%/10$^3\AA$)
which is reasonably due to the higher airmasses involved.
The obtained flat behavior makes us confident about the reliability of our results.
As for the first of our observing nights (17-18/12/2012), we do not have other solar analogs
to compare with that used to obtain the Ceres' reflectance; however, Fig.~\ref{ratio_all}
shows that its division by a solar analog observed the following night still presents a flat spectrum;
the same result is found when dividing the spectra of two solar analog stars observed during the two
different nights we had in January 2013, as well as when dividing two solar analog stars whose spectra
where acquired in December 2012 and January 2013.
All of these ratios show some telluric features, which are of course natural
because of the different observing conditions
(this is why all of the Ceres' spectra have been calibrated using a star observed close not only in airmass
but also in time)
and do not affect the overall spectral slope.
What above strongly suggest that the slope variability we present in this work
is indeed due to some short-term processes acting over the surface of Ceres.

\end{appendix}

\begin{appendix} 
\section{About a possible link between ours and Herschel's observations}
The likely sources of the water vapour detected by the Herschel Space Observatory
are two 60-km wide dark regions centred at long. 123\degr, lat. +21\degr and long. 231\degr, lat. +23\degr, respectively.
In this work we report a temporal spectral variability of Ceres,
for nearly equatorial observations centred at planetocentric longitudes $\sim$25$\degr$ and $\sim$270$\degr$,
and suggest this could be associated with variable amounts of exposed water ice.
One could object that it's difficult to explain why such longitudes are not coincinding,
given that ours and Herschel's observations were made in the same timeframe,
and we propose that the spectral changes could be related with outgassing episodes like those seen by Herschel.
Moreover, it would be difficult to imagine that the slope changes we measured are provoked by such small active areas.
In this regard, we stress that
i) our spectra are disk-integrated, hence it is basically impossible to state
where the water ice could exactly lie on the observed surface;
ii) the Herschel telescope detected water vapour, while we are suggesting the presence of water ice on the surface;
K\"uppers et al. (2014) suggested as likely sources of most of the evaporating water two dark regions, as these are warmer
than the average surface, resulting in efficient sublimation of possible water ice reservoirs.
The water ice we observe could be more concentrated in other regions of the surface, where dark (hotter) material is less abundant
and/or at higher latitudes where sublimation could be slower;
iii) the 60-km size estimated in K\"uppers et al. (2014) is somewhat model-dependent: varying the involved parameters
(e.g. the mixing ratio of active/inactive components and its spatial variation over the surface)
one can change the size of the ``most active'' spots;
iv) our observations performed in December 2012 show bluer spectra (i.e. more water ice following our assumption) than in January 2013;
Herschel clearly detected water in October 2012, whereas in March 2013 the emission line was much weaker.
Hence both the datasets could suggest a decreasing quantity of exposed water ice going from late 2012 to early 2013.
\end{appendix}

\begin{appendix} 
\section{About the 3-$\mu$m band variation}
Rivkin and Volquardsen (2010) saw variations of a few percent over Ceres' surface.
As the Rivkin et al. (2006) spectrum is an average over the entire rotational period of Ceres,
while the Vernazza et al. (2005) spectrum covers a smaller portion of the surface,
it may be that the spectrum didn't change at the longitude of the Vernazza observations.
While we cannot exclude such a possibility, it would imply that Rivkin et al.
(2006) were wrong in excluding the presence of water ice. Here we suggest that water ice is
indeed present on the surface of Ceres. Its superficial abundance could be
variable and concentrated in particular (colder) regions.
We hope that the Dawn mission will clarify this issue,
despite that it will arrive at Ceres post-perihelion,
when the heliocentric distance (around 2.9 AU) won't be favorable for detecting cometary-like activity.
\end{appendix}

\end{document}